\documentclass[12pt]{article}
\usepackage{amsmath,amssymb}
\usepackage[dvipdfmx]{graphicx}
\begin{document}

\title{Theory of general theta relations, 
addition formulas, and theta constants identities 
}
\author{Kiyoshi \textsc{Sogo}
\thanks{EMail: sogo@icfd.co.jp}
}
\date{}

\maketitle

\begin{center}
Institute of Computational Fluid Dynamics, 1-16-5, Haramachi, Meguroku, 
Tokyo, 152-0011, Japan
\end{center}
\abstract{
Jacobi's theta relations among quartic products of theta functions are generalized to those of arbitrary $n$ products. 
Igusa's procedure of derivation is extended to prove such general theta relations, from which we obtain general addition formulas 
and theta constants identities. To complete the proof, the concept of {\it cycle number $\lambda$} is essential. 
The case of $n=3$ is discussed and examined explicitly.
}
%
%

\section{Introduction}
\setcounter{equation}{0}

In his theory ({\it Fundamenta Nova}, 1829 \cite{Jacobi}) of elliptic theta functions 
$\vartheta_{j}(z, \tau)\ (j=1,2,3,4)$, Jacobi introduced  variables transformation
\begin{align}
(w_1,w_2,w_3,w_4)=(z_1,z_2,z_3,z_4)\ A,
\end{align}
where $4\times 4$ matrix $A$ is defined by
\begin{align}
A=\frac{1}{2}\left(
\begin{array}{cccc}
1&1&1&1\\
1&1&-1&-1\\
1&-1&1&-1\\
1&-1&-1&1\end{array}
\right),
\label{matrixA}
\end{align}
which has a remarkable property $A^2=E$, with the unit matrix $E$. 
For the sake of conveniences, we rewrite Jacobi's theta functions by
\begin{align}
\vartheta_{\alpha\beta}(z,\tau)=
\sum_{n=-\infty}^\infty {\bf e}\left[ \frac{1}{2}\left(n+\frac{\alpha}{2}\right)^2\tau+\left(n+\frac{\alpha}{2}\right)\left(z+\frac{\beta}{2}\right)\right],
\end{align}
where we set as usual ${\bf e}[*]=\text{exp}[2\pi i *]$, $\text{Im}\ \tau>0$, and $\alpha, \beta=0, 1$. 
Parameters $(\alpha, \beta)$ or $(\alpha/2, \beta/2)$ are called {\it theta characteristics}, 
which will be generalized further in later sections. 

Due to $A^2=E$, we have in general 
\begin{align}
\begin{split}
&\text{if}\quad (\eta_1,\eta_2,\eta_3,\eta_4)=(\xi_1,\xi_2,\xi_3,\xi_4)\ A, \\ 
&\text{then}\quad 
(\xi_1,\xi_2,\xi_3,\xi_4)=(\eta_1,\eta_2,\eta_3,\eta_4)\ A, \\
&\text{and}\quad
\sum_{j=1}^4 \xi_j^2=\sum_{j=1}^4\eta_j^2.
\end{split}
\end{align}
By using such properties effectively, Jacobi derived famous theta relations. One of many such relations is 
\begin{align}
\begin{split}
&2\ (00)=(00)'+(01)'+(10)'+(11)',\\
&(\alpha\beta)=\prod_{j=1}^4 \vartheta_{\alpha\beta}(z_j, \tau),\quad
(\alpha\beta)'=\prod_{j=1}^4 \vartheta_{\alpha\beta}(w_j, \tau).
\end{split}
\label{00A}
\end{align}
Such abbreviation imitates that of {\it A Course of Modern Analysis} by Whittaker and Watson \cite{WW}, although 
they used parenthesis $[\alpha\beta]$ instead of $(\alpha\beta)$. 
The symbol $[\alpha\beta]$ is reserved in this paper to express theta constant $\vartheta_{\alpha\beta}(0, \tau)$. 

The idea of present paper was inspired by a remark in \S 21.22 of the same book \cite{WW}, which says 
another transformation by matrix $S$ 
\begin{align}
\begin{split}
(y_1,y_2,y_3,y_4)=(x_1,x_2,x_3,x_4)\ S,\\
S=\frac{1}{2}\left(
\begin{array}{cccc}
-1&1&1&1\\
1&-1&1&1\\
1&1&-1&1\\
1&1&1&-1\end{array}
\right),
\end{split}
\label{matrixS}
\end{align}
has not only the properties $S^2=E$ and $\sum_{j=1}^4 x_j^2=\sum_{j=1}^4 y_j^2$, 
but also $\sum_{j=1}^4 x_j=\sum_{j=1}^4 y_j$, therefore {\it $S$ has more symmetry than $A$}. 
The matrix $S$ was originally introduced by Smith \cite{Smith} in 1866, which is cited also in \cite{WW}. 
In due course, Smith derived theta relations by using $S$ such as 
\begin{align}
\begin{split}
&2\ (00)=(00)'+(01)'+(10)'-(11)',\\
&(\alpha\beta)=\prod_{j=1}^4 \vartheta_{\alpha\beta}(x_j, \tau),\quad
(\alpha\beta)'=\prod_{j=1}^4 \vartheta_{\alpha\beta}(y_j, \tau).
\end{split}
\label{00S}
\end{align}
Readers are invited to compare \eqref{00S} with \eqref{00A}. The coefficients are given by 
${\bf e}(-\alpha\beta/2)$ in \eqref{00S}, that is
\begin{align}
2\ (00)=\sum_{\alpha,\beta=0,1} {\bf e}(-\alpha\beta/2)\cdot (\alpha\beta)',
\label{correct}
\end{align}
which gives the last coefficient in the right hand side of \eqref{00S} by ${\bf e}\left(-1/2\right)=e^{-\pi i}=-1$. 

If we set $x_1=x_2=x_3=x_4\equiv x$, which gives $y_1=y_2=y_3=y_4=x$ also, then \eqref{00S} becomes
\begin{align}
&(00)=(01)+(10)-(11),\nonumber \\
&\Longleftrightarrow\quad\vartheta_{00}^4(x,\tau)+\vartheta_{11}^4(x,\tau)=\vartheta_{01}^4(x,\tau)+\vartheta_{10}^4(x,\tau)
\label{JacobiQuartic}
\end{align}
which is the famous quartic theta functions identity by Jacobi \cite{Jacobi, WW}. 

A few years ago, the author happened to find a transformation by $3\times 3$ matrix $S_3$
\begin{align}
\begin{split}
(\eta_1,\eta_2,\eta_3)=(\xi_1,\xi_2,\xi_3)\ S_3, \\
S_3=\frac{1}{3}\left(
\begin{array}{ccc}
-1&2&2\\
2&-1&2\\
2&2&-1\end{array}
\right),
\end{split}
\end{align}
has the similar properties $S_3^2=E,\ \sum_{j=1}^3\xi_j^2=\sum_{j=1}^3\eta_j^2$, and 
$\sum_{j=1}^3\xi_j=\sum_{j=1}^3 \eta_j$. 
While doing some efforts to find a rule of coefficients for theta relations among products of {\it three} theta functions, 
the author again found general $n\times n$ matrix
\begin{align}
S_n=\frac{1}{n}\left(\begin{array}{ccccc}
2-n&2&\cdots&2&2\\
2&2-n&\cdots&2&2\\
\cdots&\cdots&\cdots&\cdots&\cdots\\
2&2&\cdots&2&2-n\end{array}
\right)
\equiv\frac{2}{n}\ \Lambda-E,
\end{align}
where $\Lambda$ is the matrix whose elements are all one, and $E$ is the unit matrix. 
Since $\Lambda^2=n\Lambda$, it is easy to prove $S_n^2=E$. 
Obviously Smith's $S$ is our $S_4$ and the above $S_3$ is the case of $n=3$. 
These facts have encouraged the author to construct general theory of theta relations, and the present paper is the first report of 
such efforts. 

The purpose of the present paper is to derive the coefficients of general theta relations, and to discuss addition formulas 
and theta constants identities. In the next section we will employ the procedure of Igusa's book \cite{Igusa} (chapter IV, \S 1), 
however his method is based on the matrix $A$, and is restricted to $n=4$. Hence it must be modified here to be based on 
the matrix $S_n$, and to be extended to products of arbitrary $n$ theta functions, and that is done in this paper. 

\section{General theta relations}
\setcounter{equation}{0}

\subsection{Cycle number and halving phenomena}
Let us begin by extending the definition of $S_n$ for cases of arbitrary genus $g\geq 1$,
\begin{align}
&S_n=\frac{2}{n}\ \Lambda-E=
\frac{2}{n}\left(\begin{array}{ccc}
1&\cdots&1\\
\cdots&\cdots&\cdots\\
1&\cdots&1\end{array}\right)-
\left(\begin{array}{ccc}
1&\cdots&0\\
\cdots&\cdots&\cdots\\
0&\cdots&1\end{array}\right),\nonumber \\ 
&\text{all elements}\ 1=1_g\ \text{or}\ 1_{2g},
\label{GeneralS}
\end{align}
where symbol $1_m$ means $m\times m$ unit matrix (see Igusa \cite{Igusa} p.137). 

Before discussing theta relations derived from $S_n$, let us introduce a parameter $\lambda$, 
which we call {\it cycle number}, according to a similarity to the cyclic group, defined by
\begin{align}
\lambda= 
\begin{cases}
n\quad &(n=\text{odd}),\\
\frac{n}{2}\quad &(n=\text{even}),
\end{cases}
\end{align}
which can be expressed by the following table of $\lambda$ for $n\geq 3$.

\begin{center}
\begin{tabular}{|l|ccccccccc|}
\hline
$n$& $3$ & $4$ & $5$ & $6$ & $7$ & $8$ & $9$ & $10$ & $\cdots$\\
\hline
$\lambda$ & $3$ & $2$ & $5$ & $3$ & $7$ & $4$ & $9$ & $5$ & $\cdots$\\
\hline
\end{tabular}
\end{center}
We may call such phenomena {\it halving} for the cases of $n$ even. 

The reason of introducing cycle number $\lambda$ is as follows. Let us suppose $\xi_1,\cdots,\xi_n$ are integers, and 
consider transformation
\begin{align}
(\eta_1,\cdots,\eta_n)=(\xi_1,\cdots,\xi_n)\ S_n,
\end{align}
then each $\eta_j$ has a form of ${\bf Z}+\frac{\ell}{\lambda}$ ($\ell=0, 1, \cdots, \lambda-1$). 
For any pair of $(j,\ k)$, we further find
\begin{align}
\eta_j-\eta_k=\xi_k-\xi_j\in{\bf Z},
\end{align}
which implies that $\eta_j$ and $\eta_k$ have a common form  of ${\bf Z}+\frac{\ell}{\lambda}$ with the same $\ell$. 
The value of $\ell$ is determined by the combinations of $(\xi_1,\cdots,\xi_n)\in {\bf Z}^{n}$. 
Such situation we call that all $\eta_j$'s belong to the same class $[\ell]_\lambda$, 
\begin{align}
[\ell]_\lambda=\text{the set of\ }\left\{ {\bf Z}+\frac{\ell}{\lambda}\right\},\qquad (\ell=0,\ 1,\ \cdots,\ \lambda-1)
\end{align}
or symbol $[\ell]$ can be used as simpler notation. 

\subsection{Theta functions and theta characteristics}
Let us introduce new notation for theta functions following Igusa \cite{Igusa}, 
\begin{align}
\vartheta_\mu(z,\tau)=
\sum_{\xi\in {\bf Z}^g}{\bf e}\left[\frac{1}{2}(\xi+\mu')\tau\ {}^{t}(\xi+\mu')+(\xi+\mu')\ {}^{t}(z+\mu'')\right],
\end{align}
where ${}^{t}(\xi+\mu')$ is column vector, the transpose of row vector $\xi+\mu'$, 
and $\tau$ is $g\times g$ Riemann's period matrix with the genus $g$. 
We will use sometimes another symbol $a \cdot b$ for the inner-product between two $g$-vectors, such that
\begin{align}
\begin{split}
&a=(a_1,\cdots,a_g),\quad b=(b_1,\cdots,b_g),\qquad
a\ {}^{t}b=a\cdot b,\\
&\text{and}\quad (\xi+\mu')\tau\ {}^{t}(\xi+\mu')=(\xi+\mu')\tau\cdot(\xi+\mu').
\end{split}
\end{align}

The parameters $\mu$ is called {\it theta characteristics}
\begin{align}
\mu=\binom{\mu'}{\mu''}=\binom{\mu_1'\cdots \mu_g'}{\mu_1''\cdots\mu_g''},
\label{ThetaChar}
\end{align}
where each $\mu_\alpha',\ \mu_\alpha'',\ (\alpha=1,\cdots,g)$ takes value one of 
$0/\lambda, 1/\lambda,\cdots,(\lambda-1)/\lambda$ in standard form. 
Such theta characteristics may be written by
\begin{align}
\binom{\mu'}{\mu''}=\binom{m'}{m''}_\lambda,\qquad 
\text{if}\quad \mu'=\frac{m'}{\lambda},\quad \mu''=\frac{m''}{\lambda},
\end{align}
avoiding fractions such as
\begin{align}
\binom{0}{\frac{1}{2}}=\binom{0}{1}_2,\quad 
\binom{\frac{1}{3}\frac{2}{3}}{\frac{1}{3}0}=\binom{12}{10}_3.
\end{align}
We will sometimes write theta function $\vartheta_\mu(z, \tau)$ by simpler forms 
\begin{align}
\vartheta_\mu(z, \tau)=\vartheta\binom{\mu'}{\mu''}(z, \tau)=\vartheta\binom{\mu'}{\mu''}=
\binom{\mu'}{\mu''}(z,\tau)=\binom{\mu'}{\mu''}.
\label{ThetaSymbol}
\end{align}
The last form however might be too simplified and confusing.

\subsection{Theta relations}

By paraphrasing Igusa's procedure in \cite{Igusa} pp.136-139, we can derive the following lemma and theorem. 
Let us start to write our lemma.\\

\noindent
{\bf (Lemma)}\ If we define for genus $g$, by using $S_n$ of \eqref{GeneralS},
\begin{align}
L_1={\bf Z}^{ng},\quad
L_2={\bf Z}^{ng}\ S_n,\quad
L=L_1+L_2={\bf Z}^{ng}+{\bf Z}^{ng}\ S_n,
\end{align}
the following identity holds
\begin{align}
[L: L_1]\cdot\sum_{\xi\in L_1}\Phi(\xi)=\sum_{\chi,\zeta}\left(\sum_{\eta\in L_2}\chi(\eta+\zeta)\Phi(\eta+\zeta)\right),
\label{Lemma}
\end{align}
where $[L: L_1]=[L: L_2]=\lambda^g$, with the cycle number $\lambda$ introduced in \S 2.1. 
This lemma can be proved similarly as \cite{Igusa}. 

Here we specialize $\Phi(\xi)$ by setting 
\begin{align}
\Phi(\xi)=\prod_{j=1}^n {\bf e}\left[ \frac{1}{2}(\xi_j+\mu_j')\tau\cdot(\xi_j+\mu_j')+
(\xi_j+\mu_j')\cdot(z_j+\mu_j'') \right],
\end{align}
for every $\xi=(\xi_1,\cdots,\xi_n)$ in $L$, where each $\xi_j$'s are $g$-vectors. This gives the left hand side (LHS) of 
\eqref{Lemma} by
\begin{align}
\text{LHS}=\lambda^g \cdot \sum_{\xi\in L_1} \Phi(\xi)=\lambda^g\cdot\prod_{j=1}^n \vartheta\binom{\mu_j'}{\mu_j''}(z_j, \tau).
\end{align}

Then we consider the transformations
\begin{align}
\begin{split}
&(w_1,\cdots,w_n)=(z_1,\cdots,z_n)\ S_n,\\
&(\nu_1,\cdots,\nu_n)=(\mu_1,\cdots,\mu_n)\ S_n,
\end{split}
\label{mu=nu}
\end{align}
and put
\begin{align}
\chi(\xi)={\bf e}\left[ \sum_{j=1}^n \xi_j\cdot a'' \right],
\end{align}
where we take $a''$ in $\frac{1}{\lambda}{\bf Z}^g$. If $a''$ runs over a complete set of representatives of 
$\frac{1}{\lambda}{\bf Z}^g/{\bf Z}^g$, $\chi$ runs over the dual of $L/L_1$. 
If $a'$ runs over a complete set of representatives of $\frac{1}{\lambda}{\bf Z}^g/{\bf Z}^g$, then $\zeta=(a',\cdots, a')$ 
runs over a complete set of representatives of $L/L_2$. 

Now let us compute the right hand side (RHS) of \eqref{Lemma}. 
If we put $\xi=(p_1,\cdots,p_n),\ \eta=(q_1,\cdots, q_n)$, 
then $\eta+\zeta=(q_1+a', \cdots, q_n+a')$, which we write $(r_1,\cdots, r_n)$. 
Then RHS becomes
\begin{align}
&\text{RHS}=\sum_a \sum_{\eta\in L_2}{\bf e}\left[ \sum_{j=1}^n q_j\cdot a''\right] \times
\nonumber \\
&\times {\bf e}\left[\frac{1}{2}\sum_{j=1}^n (r_j+\mu_j')\tau\cdot(r_j+\mu_j')+
\sum_{j=1}^n (r_j+\mu_j')\cdot(z_j+\mu_j'')\right].
\end{align}
By the definition of $L_2$, $(p_1,\cdots,p_n)=(q_1,\cdots,q_n)\ S_n$ runs over ${\bf Z}^{ng}$, and the equality
\begin{align}
&(r_1,\cdots,r_n)=(q_1+a',\cdots,q_n+a')=(p_1+a',\cdots,p_n+a')\ S_n
\nonumber \\
&\Longleftrightarrow\quad
(r_1,\cdots,r_n)\ S_n=(p_1+a',\cdots,p_n+a')
\end{align}
holds. 
Therefore we can rewrite
\begin{align}
\begin{split}
&\sum_{j=1}^n {}^t (r_j+\mu_j')(r_j+\mu_j')=
\sum_{k=1}^n{}^t(p_k+a'+\nu_k')(p_k+a'+\nu_k'),\\
&\sum_{j=1}^n (r_j+\mu_j')\ {}^t (z_j+\mu_j'')=\sum_{k=1}^n (p_k+a'+\nu_k')\ {}^t (w_k+\nu_k'').
\end{split}
\end{align}
And therefore the right hand side becomes 
\begin{align}
\text{RHS}&=\sum_{a}\sum_{p_j\in{\bf Z}^g} {\bf e}\left[ \sum_{j=1}^n q_j\cdot a''\right] \times 
\nonumber \\
&\times {\bf e}\left[ \frac{1}{2}\sum_{j=1}^n (p_j+\nu_j'+a')\tau\ {}^t(p_j+\nu_j'+a')\right]\times
\nonumber \\
&\times {\bf e}\left[\sum_{j=1}^n (p_j+\nu_j'+a')\ {}^t(w_j+\nu_j'')\right].
\label{2_19}
\end{align}
Since $w_j+\nu_j''=w_j+\nu_j''+a''-a''$, \eqref{2_19} can be rewritten as
\begin{align}
&\sum_a \sum_{p_j\in{\bf Z}^g} {\bf e}\left[ \sum_{j=1}^n q_j\cdot a''-\sum_{j=1}^n\left(p_j+\nu_j'+a'\right)\cdot a''\right]
\times
\nonumber \\
&\times{\bf e}\left[ \sum_{j=1}^n \frac{1}{2}(p_j+\nu_j'+a')\tau\cdot(p_j+\nu_j'+a')\right]\times
\nonumber \\
&\times {\bf e}\left[\sum_{j=1}^n (p_j+\nu_j'+a')\cdot(w_j+\nu_j''+a'')\right],
\end{align}
where due to the conservation law $\sum_{j=1}^n q_j=\sum_{j=1}^n p_j$, we obtain the coefficient 
\begin{align}
&{\bf e}\left[ \sum_{j=1}^n q_j\cdot a''-\sum_{j=1}^n\left(p_j+\nu_j'+a'\right)\cdot a''\right]
\nonumber \\
&\quad ={\bf e}\left[ -\sum_{j=1}^n (\nu_j'+a')\cdot a'' \right].
\label{tobehalved}
\end{align}
Thus finally we get the following theorem.\\

\noindent
{\bf (Theorem)}\quad Under the transformations
\begin{align}
\begin{split}
&(w_1,\cdots,w_n)=(z_1,\cdots,z_n)\ S_n,\\
&(\nu_1,\cdots,\nu_n)=(\mu_1,\cdots,\mu_n)\ S_n,
\end{split}
\label{Mu=Nu}
\end{align}
we have the theta relations
\begin{align}
&\lambda^g \cdot \left\{\vartheta_{\mu_1}(z_1,\tau)\cdots\vartheta_{\mu_n}(z_n,\tau)\right\}
\nonumber \\
&\qquad=\sum_a {\bf e}\left[-\sum_{j=1}^n (\mu_j'+ a')\cdot a''\right]\times \nonumber \\
&\qquad\qquad\quad\times \left\{ \vartheta_{\nu_1+a}(w_1, \tau)\cdots \vartheta_{\nu_n+a}(w_n, \tau) \right\},
\label{theorem}
\end{align}
where we have replaced $\sum_j \nu_j'$ by $\sum_j \mu_j'$, since $S_n$ allows the equalities
\begin{align}
\sum_{j=1}^n \mu_j'=\sum_{j=1}^n \nu_j',\quad \sum_{j=1}^n \mu_j''=\sum_{j=1}^n \nu_j'',
\end{align}
which are also one of the conservation laws of \eqref{Mu=Nu}.\\

This theorem looks good at least for {\it odd} $n$, but there remains something wrong for {\it even} $n$, 
which can be checked explicitly by examining the following corollary. \\

\noindent
{\bf (Corollary)}\quad For the simplest case of $\mu_1=\cdots=\mu_n=\binom{0}{0}$, the theorem becomes
\begin{align}
\lambda^g\cdot\prod_{j=1}^n \vartheta\binom{0}{0}(z_j, \tau)=
\sum_a {\bf e}\left(-n a'\cdot a''\right)\cdot
\prod_{j=1}^n \vartheta\binom{a'}{a''}(w_j, \tau),
\label{corollary}
\end{align}
because $\nu_1=\cdots=\nu_n=\binom{0}{0}$ also.\\

This formula \eqref{corollary} looks again good for odd $n$, 
but for even $n$ it has an exceptional {\it error} at least for the case of $n=4\ (\lambda=2),\ g=1$ 
(the case of Smith's theta functions), because the coefficient is
\begin{align}
\text{not}\quad {\bf e}\left(-2 a'a''\right),\quad \text{but}\quad {\bf e}\left(-4 a'a''\right),
\end{align}
although the former is the correct one, since $2a'a''=\alpha\beta/2$ of \eqref{correct}. 
The value of the latter formula gives always $=1$, and is incorrect. 

In this way, it seems that the coefficient in \eqref{corollary} should be 
\begin{align}
{\bf e}\left(-\lambda a'\cdot a''\right)\quad\text{instead of}\quad {\bf e}\left(-n a'\cdot a''\right),
\end{align}
which gives no change for odd $n$ because $\lambda=n$.\\

Such prescription $n\rightarrow\lambda$ can be understood by the following discussions. 
The transformation, $(\nu_1,\cdots,\nu_n)=(\mu_1,\cdots,\mu_n)\ S_n$, 
can be rewritten by using the conserved quantity $s\equiv\lambda^{-1}\sum_j\mu_j=\lambda^{-1}\sum_j \nu_j$, as
\begin{align}
\nu_j=s-\mu_j,
\end{align}
which was already noticed by Smith \cite{Smith} for $n=4$ case. 

When $n$ is {\it even}, therefore, we can divide the theta characteristics in two groups of symmetric and anti-symmetric ones, 
denoted by $(j, j+1)$'s and $[j, j+1]$'s, defined by
\begin{align}
(j, j+1)=\frac{1}{2}\left(\mu_j+\mu_{j+1}\right),\quad
[j, j+1]=\frac{1}{2}\left(\mu_j-\mu_{j+1}\right),
\end{align}
for $j=1,3,\cdots,n-1$.

Then it is obvious that the transformations among each $(j, j+1)$'s and $[j, j+1]$'s are independent and decoupled.
In other words, when $n$ is even the transformations $(\nu_1,\cdots,\nu_n)=(\mu_1,\cdots,\mu_n)\ S_n$ are replaced and 
{\it halved}, by the transformations
\begin{align}
\begin{split}
&\frac{1}{2}\left(\nu_j+\nu_{j+1}\right)=s-\frac{1}{2}\left(\mu_j+\mu_{j+1}\right),\\
&\frac{1}{2}\left(\nu_j-\nu_{j+1}\right)=0-\frac{1}{2}\left(\mu_j-\mu_{j+1}\right),
\end{split}
\end{align}
among $\lambda\ (=n/2)$ quantities of $j=1,3,\cdots,n-1$.

If the above discussions for even $n$ are considered in the computations of RHS of \eqref{Lemma}, we recognize that 
the prescription $n\rightarrow\lambda$ is indeed justified. 
Thus the coefficient of \eqref{theorem} should be replaced by
\begin{align}
\begin{split}
&{\bf e}\left[ -\left( \sum_{j=1}^n \mu_j'+\lambda a'\right)\cdot a'' \right],
\end{split}
\label{haived}
\end{align}
which can be verified by other examples of theta relations given in \cite{WW} p.468. 
In conclusion, the theorem \eqref{theorem} must be modified as follows.\\

\noindent
{\bf (Theorem modified)}\quad For arbitrary $n$, we have
\begin{align}
&\lambda^g \cdot \left\{\vartheta_{\mu_1}(z_1,\tau)\cdots\vartheta_{\mu_n}(z_n,\tau)\right\}
\nonumber \\
&\qquad =\sum_a {\bf e}\left[-(\sum_{j=1}^n \mu_j'+ \lambda a')\cdot a''\right]\times \nonumber \\
&\qquad\qquad\quad\times \left\{ \vartheta_{\nu_1+a}(w_1, \tau)\cdots \vartheta_{\nu_n+a}(w_n, \tau) \right\}.
\label{theorem_modified}
\end{align}
And the corollary \eqref{corollary} must be also modified as follows.\\

\noindent
{\bf (Corollary modified)}\quad For arbitrary $n$, 
\begin{align}
\lambda^g\cdot\prod_{j=1}^n \vartheta\binom{0}{0}(z_j, \tau)=
\sum_a {\bf e}\left(-\lambda a'\cdot a''\right)\cdot
\prod_{j=1}^n \vartheta\binom{a'}{a''}(w_j, \tau).
\label{halved}
\end{align}
Therefore, {\it halving} phenomena of the coefficients also occur for $n$ even.\\

It should be mentioned here that we can verify the same result, in another way, by considering a limit of Riemann's 
period matrix $\tau\rightarrow 0$, 
which makes situation simpler and writes the result by using Dirac's delta functions, 
because theta functions become infinite sums of exponential functions in such limit.

\section{Addition formulas and theta constants identity}
\setcounter{equation}{0}
\subsection{Addition formulas}

Let us consider the case of $n=3$ in detail, which implies $\lambda=3$ also. 
For arbitrary genus $g$, from \eqref{halved} we have theta relation
\begin{align}
3^g\cdot \binom{0}{0}=\sum_a\ {\bf e}\left(-3 a'\cdot a''\right)\cdot \binom{a'}{a''}',
\end{align}
where we used abbreviation
\begin{align}
&\binom{\mu'}{\mu''}=\prod_{j=1}^3 \vartheta_{\mu}(x_j, \tau),\quad
\binom{\nu'}{\nu''}'=\prod_{j=1}^3 \vartheta_{\nu}(y_j, \tau),\\
&(y_1,\ y_2,\ y_3)=(x_1,\ x_2,\ x_3)\ S_3.
\end{align}
Since we are considering the case of $\mu=\nu=0$ in \eqref{theorem_modified}, only $a=\binom{a'}{a''}$ is remaining commonly.

The variables transformation is given explicitly by
\begin{align}
\begin{split}
&y_1=\frac{1}{3}\left(-x_1+2x_2+2x_3\right),\\
&y_2=\frac{1}{3}\left(2x_1-x_2+2x_3\right),\\
&y_3=\frac{1}{3}\left( 2x_1+2x_2-x_3\right),
\end{split}
\end{align}
where $x_j$'s and $y_j$'s are $g$-vectors, therefore the situations are still complicated for general $g$. 
Note that theta characteristics $a'$ and $a''$ are also $g$-vectors. 

For the case of $g=1$, the result is given explicitly by
\begin{align}
\begin{split}
3\cdot\binom{0}{0}&=
\binom{0}{0}'+\binom{0}{\frac{1}{3}}'+\binom{0}{\frac{2}{3}}'
+\binom{\frac{1}{3}}{0}'+\omega^2 \binom{\frac{1}{3}}{\frac{1}{3}}'+\omega\binom{\frac{1}{3}}{\frac{2}{3}}'\\
&\quad+\binom{\frac{2}{3}}{0}'+\omega \binom{\frac{2}{3}}{\frac{1}{3}}'+\omega^2\binom{\frac{2}{3}}{\frac{2}{3}}',
\qquad (\omega=e^{2\pi i/3})
\end{split}
\label{3_1}
\end{align}
where the right hand side is a sum of nine (=$3^2$) terms. \\
Similarly for the case of $g=2$, the theta relation is given by
\begin{align}
3^2\cdot\binom{00}{00}=\text{a sum of eighty-one ($=9^2$) terms},\ 
\label{3_2}
\end{align}
with coefficients one of $(1,\ \omega,\ \omega^2)$. Such formulas are the theta relations of $n=3$. 

Theta relations should also be called the {\it addition formulas}, 
which express three product of theta functions of variables $x_j$'s 
by those of functions of $y_j$'s, related by $(y_1, y_2, y_3)=(x_1, x_2, x_3) S_3$. 
From such addition formula, we can obtain {\it ternary theta functions identity} by setting $x_1=x_2=x_3\equiv x$, 
and $y_1=y_2=y_3=x$ as a consequence, which is an analogue of quartic theta functions identity \eqref{JacobiQuartic} of $n=4$ 
found by Jacobi. 
For an example, from \eqref{3_1} of $g=1$, we obtain
\begin{align}
3\cdot\vartheta\binom{0}{0}^3&=\vartheta\binom{0}{0}^3+\vartheta\binom{0}{\frac{1}{3}}^3+\vartheta\binom{0}{\frac{2}{3}}^3+
\vartheta\binom{\frac{1}{3}}{0}^3+\vartheta\binom{\frac{2}{3}}{0}^3
\nonumber \\
&+\omega\ \vartheta\binom{\frac{1}{3}}{\frac{2}{3}}^3+\omega\ \vartheta\binom{\frac{2}{3}}{\frac{1}{3}}^3+
\omega^2\ \vartheta\binom{\frac{1}{3}}{\frac{1}{3}}^3+\omega^2\ \vartheta\binom{\frac{2}{3}}{\frac{2}{3}}^3,
\end{align}
where we used the abbreviation
\begin{align}
\vartheta\binom{\mu'}{\mu''}^3=\vartheta\binom{\mu'}{\mu''}^3(x,\tau).
\end{align}


\subsection{Theta constants identities}

Theta constants identities are equalities derived from theta relations by substituting $x_1=x_2=x_3=0$ and $y_1=y_2=y_3=0$ as 
a result. For the case of $n=3,\ g=1$, we have from the theta relation \eqref{3_1} 
\begin{align}
3\cdot\genfrac{[}{]}{0pt}{}{0}{0}^3&=
\genfrac{[}{]}{0pt}{}{0}{0}^3+\genfrac{[}{]}{0pt}{}{0}{\frac{1}{3}}^3+\genfrac{[}{]}{0pt}{}{0}{\frac{2}{3}}^3+
\genfrac{[}{]}{0pt}{}{\frac{1}{3}}{0}^3+\omega^2\genfrac{[}{]}{0pt}{}{\frac{1}{3}}{\frac{1}{3}}^3+
\omega\genfrac{[}{]}{0pt}{}{\frac{1}{3}}{\frac{2}{3}}^3
\nonumber \\
&+\genfrac{[}{]}{0pt}{}{\frac{2}{3}}{0}^3+\omega\genfrac{[}{]}{0pt}{}{\frac{2}{3}}{\frac{1}{3}}^3+
\omega^2\genfrac{[}{]}{0pt}{}{\frac{2}{3}}{\frac{2}{3}}^3,\quad (\omega=e^{2\pi i/3})
\end{align}
where we used abbreviation
\begin{align}
\genfrac{[}{]}{0pt}{}{\alpha}{\beta}=\vartheta\binom{\alpha}{\beta}(0, \tau)
=\sum_{n=-\infty}^\infty {\bf e}\left[\frac{1}{2}(n+\alpha)\tau+(n+\alpha)\beta\right].
\end{align}
At this point, we can derive the equalities
\begin{align}
\genfrac{[}{]}{0pt}{}{1-\alpha}{0}&=
\sum_{n=-\infty}^\infty {\bf e}\left[ \frac{1}{2}(n+1-\alpha)^2\tau+(n+1-\alpha)\cdot 0\right],\quad (n+1\rightarrow -n)
\nonumber \\
&=\sum_{n=-\infty}^\infty {\bf e}\left[ \frac{1}{2}(n+\alpha)^2\tau\right]
=\genfrac{[}{]}{0pt}{}{\alpha}{0},
\\
\genfrac{[}{]}{0pt}{}{0}{1-\beta}&=
\sum_{n=-\infty}^\infty {\bf e}\left[ \frac{1}{2}n^2\tau+n\cdot (1-\beta)\right],\quad (n\rightarrow -n)
\nonumber \\
&=\sum_{n=-\infty}^\infty {\bf e}\left[ \frac{1}{2}n^2\tau+n\beta-n\right]
=\genfrac{[}{]}{0pt}{}{0}{\beta},
\end{align}
where we used ${\bf e}(-n)=e^{-2\pi in}=1$, and also
\begin{align}
\genfrac{[}{]}{0pt}{}{1-\alpha}{\beta}&=
\sum_{n=-\infty}^\infty {\bf e}\left[ \frac{1}{2}(n+1-\alpha)^2\tau+(n+1-\alpha)\beta\right],
\quad (n+1\rightarrow -n)
\nonumber \\
&=\sum_{n=-\infty}^\infty {\bf e}\left[ \frac{1}{2}(n+\alpha)^2\tau-(n+\alpha)\beta\right]
\nonumber \\
&=\sum_{n=-\infty}^\infty {\bf e}\left[ \frac{1}{2}(n+\alpha)^2\tau+(n+\alpha)(1-\beta)-n-\alpha\right]
\nonumber \\
&=e^{-2\pi i\alpha}\cdot\genfrac{[}{]}{0pt}{}{\alpha}{1-\beta}.
\end{align}
Therefore we have equalities
\begin{align}
\begin{split}
&\genfrac{[}{]}{0pt}{}{\frac{1}{3}}{0}^3=\genfrac{[}{]}{0pt}{}{\frac{2}{3}}{0}^3,\quad
\genfrac{[}{]}{0pt}{}{0}{\frac{1}{3}}^3=\genfrac{[}{]}{0pt}{}{0}{\frac{2}{3}}^3,\\
&\genfrac{[}{]}{0pt}{}{\frac{1}{3}}{\frac{1}{3}}^3=\genfrac{[}{]}{0pt}{}{\frac{2}{3}}{\frac{2}{3}}^3,\quad
\genfrac{[}{]}{0pt}{}{\frac{2}{3}}{\frac{1}{3}}^3=\genfrac{[}{]}{0pt}{}{\frac{1}{3}}{\frac{2}{3}}^3.
\end{split}
\end{align}
By using these equalities, we finally obtain
\begin{align}
\genfrac{[}{]}{0pt}{}{0}{0}^3=\genfrac{[}{]}{0pt}{}{0}{\frac{1}{3}}^3+\genfrac{[}{]}{0pt}{}{\frac{1}{3}}{0}^3
+\omega \genfrac{[}{]}{0pt}{}{\frac{1}{3}}{\frac{2}{3}}^3
+\omega^2 \genfrac{[}{]}{0pt}{}{\frac{1}{3}}{\frac{1}{3}}^3,\quad (\omega=e^{2\pi i/3})
\end{align}
which is {\it ternary theta constants identity} of $n=3,\ g=1$, corresponding to Jacobi's quartic theta constants identity 
of $n=4,\ g=1$,  
\begin{align}
\genfrac{[}{]}{0pt}{}{0}{0}^4=\genfrac{[}{]}{0pt}{}{0}{\frac{1}{2}}^4+\genfrac{[}{]}{0pt}{}{\frac{1}{2}}{0}^4
\ \Longleftrightarrow\ 
\vartheta_{00}^4(0,\tau)=\vartheta_{01}^4(0,\tau)+\vartheta_{10}^4(0,\tau),
\end{align}
which is \eqref{JacobiQuartic} of $x=0$, because of $\vartheta_{11}(0,\tau)=0$. \\

\section{Summary}
\setcounter{equation}{0}

General theory of theta relations is formulated, where the important concept of {\it cycle number $\lambda$} is introduced. 
Our main results are expressed by the theorem modified \eqref{theorem_modified} 
and its corollary modified \eqref{halved}. 
They can be regarded also as the addition formulas, which bring theta functions identities and theta constants identities. 
In this first report the case of $n=3$ is discussed in details and explicitly. 

In the forthcoming second report, further applications will be discussed such as the differential equations satisfied 
by extended theta functions, and theta constants identities with general theta characteristics. 
The latter problems are deeply connected with and will simplify some of the results by Farkas and Kra \cite{FK}.

\end{document}